\documentclass[aps,prb,twocolumn,longbibliography,superscriptaddress]{revtex4-2}%
\usepackage{graphicx}
\usepackage{amssymb}
\usepackage{amsmath}
\usepackage{epsfig}
\usepackage{color}
\usepackage{mathtools}
\usepackage[colorlinks,linkcolor=blue,anchorcolor=blue,citecolor=blue,urlcolor=blue]{hyperref}
\usepackage{physics}
\usepackage{siunitx}
\usepackage{bm}
\usepackage{comment}
\usepackage[sort&compress]{natbib}
\begin{document}
\title{Orbital and spin bilinear magnetotransport effect in Weyl/Dirac semimetal}
\author{Zhanyunxin Du}
\affiliation{Department of Physics, City University of Hong Kong, Kowloon, Hong Kong SAR}
\author{Yue-Xin Huang}
\affiliation{Department of Physics, City University of Hong Kong, Kowloon, Hong Kong SAR}
\affiliation{School of Sciences, Great Bay University, Dongguan 523000, China}
\affiliation{Great Bay Institute for Advanced Study, Dongguan 523000, China}
\author{Xiao Li}
\email{xiao.li@cityu.edu.hk}
\affiliation{Department of Physics, City University of Hong Kong, Kowloon, Hong Kong SAR}
\date{\today }

\begin{abstract}
    We theoretically investigate the bilinear current, scaling as $j\sim EB$, in two- and three-dimensional systems. Based on the extended semiclassical theory, we develop a unified theory including both longitudinal and transverse currents. We classify all contributions according to their different scaling relations with the relaxation time. We reveal the distinct contributions to the ordinary Hall effect, planar Hall effect, and magnetoresistance. We further report an intrinsic ordinary Hall current, which has a geometric origin and has not been discussed previously. Our theory is explicitly applied to studying a massive Dirac model and a $\mathcal{PT}$-symmetric system. Our work presents a general theory of electric transport under a magnetic field, potentially laying the groundwork for future experimental studies or device fabrications. 
\end{abstract}

\maketitle

\section{Introduction}

Transport properties are essential phenomena in the field of condensed matter physics. 
When a sample is subject to a driving field (such as electric, magnetic, temperature gradient, or strain), a longitudinal or transverse current will be induced inside, and the corresponding transport coefficients can be measured. 
These measurements are widely used to characterize materials in electronic devices and reveal their band structures. 
Some of the notable examples include the (anomalous) Hall effect~\cite{Jungwirth2002,Onoda2002,Nagaosa2010a}, Nernst effect~\cite{Xiao2006}, giant magnetoresistance~\cite{Baibich1988,binasch_Enhanced_1989} and thermoelectric effect~\cite{ContinuumTheory}. 
These phenomena found great success in connecting theories and experiment observations, and thereby offering various routes for device applications. 

The ordinary Hall effect refers to the generation of a transverse voltage in response to a longitudinal driving current under an out-of-plane magnetic field~\cite{ashcroft_Solid_1976}.
This famous phenomenon is known to be induced by the Lorentz force. 
When the applied $B$ field lies in the transport plane, the transverse response is known as the planar Hall effect (PHE), which obviously has a distinct physical origin~\cite{Goldberg1954,Nandy2017a} and has been studied theoretically and experimentally in various systems, e.g., semimetal~\cite{Kumar2018,Nandy2017a,Ma2019,Zhang2019,Zyuzin2020,Battilomo2021}, (anti)-ferromagnets~\cite{Tang2003,Seemann2011}, and topological material~\cite{Wu2018,Cullen2021}. 
However, most previous studies on PHE focused on the regime of the extrinsic mechanisms, with contributions explicitly depending on the relaxation time $\tau$. 
Only recently, several experiments~\cite{Liang2018,Zhou2022a,Wang2024b} identified an intrinsic contribution to the PHE, i.e., a dissipationless Hall response that exhibits a linear dependence on the magnetic field. 
Moreover, some recent theoretical studies~\cite{Gao2014,Wang2024a,Xiang2024} used the extended semiclassical theory~\cite{Gao2014,Gao2015,Gao2019b} to derive various theories for the PHE. 
Such theories have been successfully applied to various nonlinear responses and obtained many important results~\cite{Wang2021a,Liu2021b,Liu2022,xiang_Thirdorder_2023a,Huang2023,Xiao2023,Xiang2024}.
In particular, for the longitudinal component, this bilinear current produces an interesting magnetoresistance contribution, which is linear in the $B$ field~\cite{Xu1997,Branford2005,Hu2008,Branford2005,Friedman2010,Zhang2011}.
This contribution differs from the quadratic magnetoresistance, which is typically insensitive to reversing the applied $B$ field. 

In this work, we present a systematic theoretical study of magneto-transport in Weyl/Dirac semimetals, focusing on the bilinear current response $j\sim EB$. 
We develop a unified theory for this second-order response based on the extended semiclassical theory and the relaxation time approximation. 
Our theory not only captures all previously proposed mechanisms but also finds important new corrections. 
We classify different contributions to the magneto-transport properties according to their dependence on the relaxation time $\tau$. 
Specifically, we find that the $\tau^0$ term not only contributes to the intrinsic PHE but also introduces a correction to the ordinary Hall effect. 
Such a correction is previously unknown and exhibits qualitatively different properties from the ordinary Hall effect induced by the Lorentz force, which is a $\tau^2$ contribution. 
Finally, the $\tau^1$ term we obtained only exists in systems with broken time-reversal ($\mathcal{T}$) symmetry and should be regarded as the $\mathcal T$-odd term. 
The advantages of our work include the following. 
First, our results are readily applicable to toy model calculations and integrable with first-principle approaches. 
Second, we can extract both the orbital and spin contributions in these effects, which are accurate to linear order in the electric and magnetic fields.

The structure of this paper is the following. 
In Section~\ref{Section:General}, we present our general theory and also break down different contributions to the magneto-transport properties according to their dependence on $\tau$. 
In Section~\ref{Section:2D} and Section~\ref{Section:3D}, we apply our theory to two representative toy models in 2D and 3D, respectively. 
These models can capture the low-energy physics in many materials. 
We carefully discuss the distinct behaviors of the different contributions and distinguish them in different regimes. 
Such a discussion also provides possible mechanisms to separate the three contributions in experiments. 
In Section~\ref{Section:Discussion}, we provide additional discussions and the conclusion. 
Our work may establish the theoretical foundation of the bilinear magneto-transport effect. 
It may also shed light on possible routes to enhance this effect in the device applications in the future. 




\begin{figure}[t]
    \centering
    \includegraphics[width=0.48\textwidth]{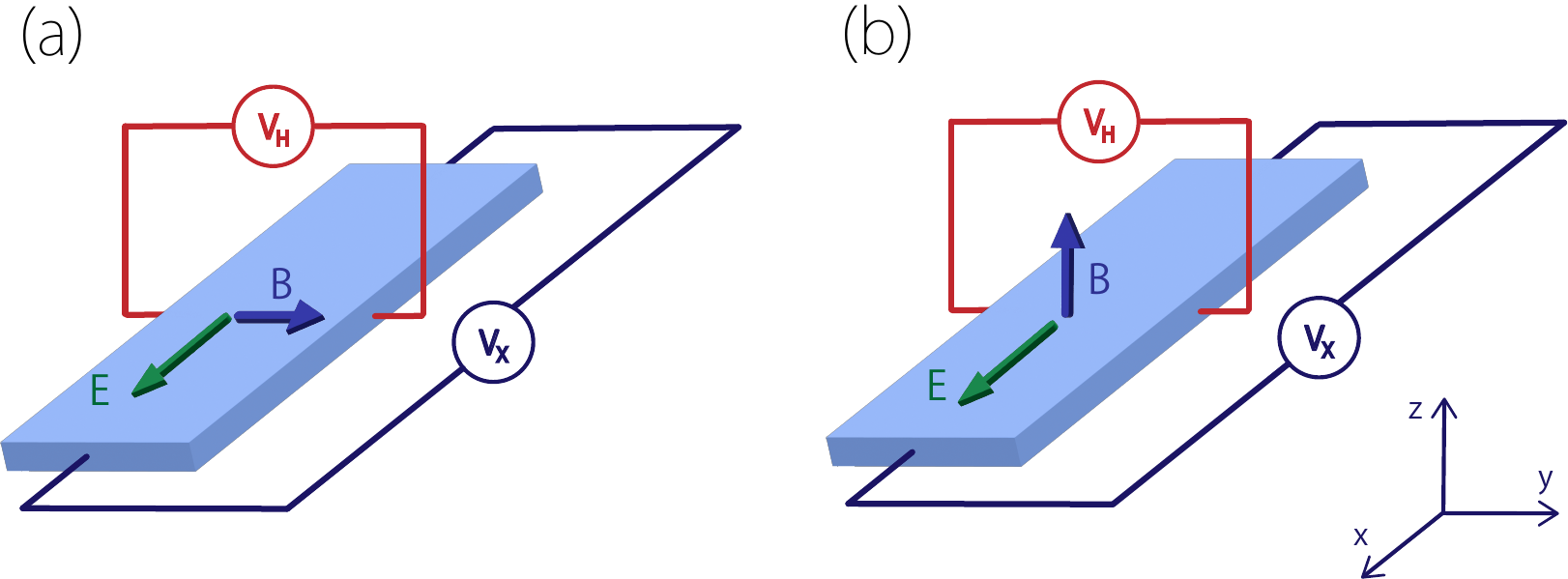}
    \caption{
        A schematic setup for a magneto-transport measurement when the magnetic field is (a) in-plane and (b) out-of-plane. 
        In (a), the transverse signal is called the planar Hall effect (PHE), as the magnetic is applied within the transport plane. In (b), the transverse signal is known as the ordinary Hall effect.
        Finally, in both cases, the longitudinal response is known as the magnetoresistance.
    }
    \label{fig-setup}
\end{figure}

\section{The general theory \label{Section:General}}
The magneto-transport can be generally divided into three different effects, as shown in Fig.~\ref{fig-setup}. 
In our discussion, we consider the $x$-$y$ plane as the transport plane. The electric field is applied within this plane, and response signals are measured in the same plane. 
In the experiment, the transport plane can be selected as the plane formed by the symmetry lines or crystal axis. 
In a non-magnetic material, a nonzero Hall effect requires breaking the time-reversal symmetry, typically achieved by applying a magnetic field. 
When the magnetic field is applied within (out of) the transport plane, the transverse signal is referred to as the planar (ordinary) Hall effect, as shown in Fig.~\ref{fig-setup}. 
Meanwhile, the longitudinal signal is called magnetoresistance. 
For simplicity, we consider the DC limit, meaning that the frequency of the driving field always stays in the $\omega \tau\ll 1$ regime.

\subsection{The formalism}
To investigate the magneto-transport effect, we look at the time evolution of a wave packet of the charged carriers as a semiclassical object and use the Boltzmann equation to describe the evolution of the wave packet distribution function in the phase space. 
Specifically, a Bloch wave packet with band index $n$ has a center at $(\vb r,\vb k)$ in phase space, then the equation of motion reads (take $e=\hbar=1$) 
\begin{align}
    \dot{\bm r}&= \pdv{\tilde\varepsilon_n}{\bm k}- \dot{\bm k}\times \tilde{\boldsymbol\Omega}_n,
    \\
    \dot{\bm k}&= -\bm E-\dot{\bm r}\times \bm B.
    \label{eq-dotrk}
\end{align}
We employed the equations from the extended semiclassical theory~\cite{Gao2014,Gao2015,Gao2019b} because the bilinear current $j\sim EB$ is a second-order response. 
These equations have the same structure as the original Chang-Sundaram-Niu equation of the conventional semiclassical theory~\cite{Chang1995,Sundaram1999,Xiao2010}. 
The only corrections are that the band energy $\tilde\varepsilon_n$ and the Berry curvature $\tilde{\boldsymbol\Omega}_n$ now include field corrections, as highlighted by the tilde in these symbols. 
In our case, such field corrections are induced by the applied magnetic field. 
As a result, the $B$ field not only serves as an external field in Eq.~\eqref{eq-dotrk}, but it also alters the band structure, thereby affecting $\tilde\varepsilon_n$ and $\tilde{\boldsymbol\Omega}_n$. 
Moreover, one should generally consider both the orbital and spin degrees of freedom that couple to the $B$ field. 
As a result, the magneto-transport effect must have two distinct contributions. 
The spin part must strongly rely on the spin-orbit coupling (SOC) strength, while the orbital part may rely on the Lorentz force or coupling with the Berry curvature. 

To calculate the desired current, we still need the distribution function of the Bloch wave packet.
Within the relaxation time approximation, the Boltzmann equation takes the following form, 
\begin{equation}
  \dot{\bm k}\cdot \nabla_{\bm k}f= -\frac{f-f_0}{\tau},
\end{equation}
where $f$ is the distribution function, and
$f_0$ is the equilibrium Fermi-Dirac distribution. 
In the semiclassical regime, the solution to the above equation can be formally written as
\begin{align}
    f=\sum_{\eta=0}^\infty (-\tau \dot{\bm k}\cdot \nabla_{\bm k})^\eta f_0(\tilde\varepsilon).
    \label{eq-fdis}
\end{align}
In general, $\dot{\vb k}$ depends on both the $B$ and $E$ field, as shown in Eq.~\eqref{eq-dotrk}.
For an in-plane setup in 2D, $\dot{\vb k}$ only depends on $E$, and the $B$ field is decoupled from the orbital degree of freedom. 
The magnetic field can only enter the conductivity from the spin degree of freedom via the correction to the energy and the Berry curvature.

With the equation of motion and the solution of the distribution function, one can then calculate the charge current as
\begin{align}
    \bm j= -\int[\dd{\bm k}]\mathcal D(\bm k)\ \dot{\bm r}f ,
    \label{eq-current}
\end{align}
where $[\dd{\bm k}]$ is a shorthand notation for $\sum_n \dd{\bm k}/(2\pi)^d$ in a $d$-dimension system, and $\mathcal D(\bm k)=1+\vb B\cdot \tilde{\boldsymbol\Omega}(\vb k)$ is a correction for the phase-space density of states~\cite{Xiao2005}. 
Finally, the explicit form of $\vb j$ can be expressed as
\begin{align}
    \vb j= -\int[\dd{\vb k}]
    \left[ \tilde{\vb v}^n+\vb E\times \tilde{\boldsymbol\Omega}_n
    +\vb B(\tilde{\vb v}^n\cdot \tilde{\boldsymbol\Omega}_n) \right]f,
    \label{eq-currentexplicit}
\end{align}
which contains three terms on the right-hand side. 
The desired bilinear current $j$ can be extracted from Eq.~(\ref{eq-currentexplicit}) by collecting the contributions that scale as $EB$.
The corresponding conductivity can then be expressed using a third-rank tensor $\chi$ as
\begin{eqnarray}
    j_a= \chi_{abc}(\tau) E_b B_c,
    \label{eq-typecurrent}
\end{eqnarray}
where the subscripts $a,b,\in \{x,y\}$ and $c\in\{x,y,z\}$ according to our setup.
Elements like $\chi_{xyz}$ or $\chi_{yxz}$ describe the ordinary Hall effect because the magnetic field is along $z$ direction.
Meanwhile, elements like $\chi_{xyy}$ capture the PHE when the magnetic field is inside the transport plane.
Finally, the longitudinal responses are classified as the magnetoresistance. 

The above is one of the central results of our work. 
Several comments are in order. 
First, the axial nature of $\vb B$ implies that the conductivity tensor $\chi$ must also be axial, consistent with Neumann's principle of a third-rank axial tensor~\cite{Powell2010}. 
Second, unlike the conductivity of the second-order Hall response, the response tensor $\chi_{abc}$ survives even in the presence of an inversion symmetry. 
Finally, the response tensor $\chi$ has both time-reversal ($\mathcal{T}$) even and odd parts~\cite{Freimuth2014,Zelezny2017,Manchon2019,Xiao2023}, which are distinguished according to their parities under $\mathcal{T}$. 
Particularly, the $\mathcal{T}$-odd part of $\chi$ can only appear in magnetic systems, whereas the $\mathcal{T}$-even part is allowed in both nonmagnetic and magnetic systems. 
Moreover, in magnetic systems, the two parts have different transformation properties under any primed operation (defined as the combination of $\mathcal{T}$ and a spatial operation). 
Clearly, the $\mathcal T$-odd or $\mathcal T$-even response depends on whether $\chi(\tau)$ is odd or even function of $\tau$. 
In the following, we shall group the contributions to $j$ (or $\chi$) according to their scaling relation with $\tau$. 
In the current theoretical framework, this leads to three types of currents: the $\tau^0$-, $\tau^1$-, and $\tau^2$-scaling currents. 

\subsection{The $\tau^0$-scaling current}
The intrinsic PHE manifests the intrinsic properties of a material, which is independent of the scattering. 
To obtain this contribution within our theoretical framework, we let $\eta=0$ in Eq.~\eqref{eq-fdis} so that only the equilibrium Fermi-Dirac distribution is involved.
The resulting current, denoted as $j^{(0)}$, has no $\tau$ dependence and hence is called the intrinsic contribution. 
The corresponding conductivity is solely determined by the band structure of the system, manifesting its significance as an intrinsic property.
The theory of this intrinsic PHE has been developed in two recent works~\cite{Gao2014,Wang2024a}, although they only focused on the case where the magnetic field lies in the transport plane (the PHE configuration). 
In this section, we briefly review those results and then extend them beyond the PHE.

To begin with, we note that in this case, only the equilibrium Fermi-Dirac distribution enters the final expression; all the field-dependent term comes from the velocity of the carrier. 
In addition, since the first-order correction of the $E$ field only shifts the wave packet and does not induce any current flow, the first term of Eq.~\eqref{eq-currentexplicit} will not contribute to $\vb j^{(0)}$. 
The second term of Eq.~\eqref{eq-currentexplicit} is the so-called anomalous velocity, contributing to the anomalous Hall effect. 
Given that we are looking for terms like $j^{(0)} \sim EB$, the remaining $B$ factor has to come from its correction to the Berry curvature $\tilde{\boldsymbol\Omega}_n$ and energy. 
This $B$ field correction takes the form of ${\bm\Omega}^B=\nabla_{\bm k}\times \bm{ \mathcal{A}}^B$, where $\mathcal{A}^B_a=F_{ab}B_b$ is the $B$-field-induced Berry connection (or positional shift of the wave packet center) and $F_{ab}$ is known as the anomalous polarizability tensor, which is a gauge invariant quantity~\cite{Gao2014}. 
As we mentioned, $F_{ab}$ should contain both spin and orbital degrees of freedom, so it can be written generally as $F_{ab}=F^O+F^S$, with anomalous spin polarizability 
\begin{align}
    F_{ab}^{S}(\vb k)&= -2\Re \sum_{m\ne n} \frac{\mathcal A^{mn}_a\mathcal M^{S,nm}_b }{\varepsilon_n-\varepsilon_m},
    \label{eq-ASP}
\end{align}
and anomalous orbit polarizability
\begin{align}
    F_{ab}^{O}(\vb k)&= -2\Re\sum_{m\ne n} \frac{\mathcal A^{mn}_a\mathcal M^{O,nm}_b}{\varepsilon_n-\varepsilon_m}
    -\frac{1}{2}\varepsilon_{bcd}\partial_c \mathcal G_{da}^n.
    \label{eq-AOP}
\end{align}
Here, $\varepsilon_{bcd}$ is the Levi-Civita symbol, $\mathcal A^{nm}_b= \ip{u_n}{i\partial_b u_m}$ is the unperturbed interband Berry connection, $\partial_b\equiv \partial_{k_b}$,
$\ket{u_n}$ is the unperturbed cell-periodic Bloch state with energy $\varepsilon_n$;
$\mathcal M^{o,mn}=\sum_{l\ne n}(\vb v^{ml}+\delta^{lm}\vb v^n)\times \mathcal A^{ln}/2$ and
$\mathcal M^{s,mn}=-g\mu_B \vb s^{mn}$ are interband orbital and spin magnetic moments, respectively;
$\mathcal G_{da}=\Re \sum_{m\ne n} \mathcal A_d^{nm}\mathcal A_a^{mn}$ is the quantum metric tensor.
The corresponding conductivity can be expressed using these quantities.
We also mention that the third term in Eq.~\eqref{eq-currentexplicit} is also known as the chiral magnetic current~\cite{Gao2019b}, which vanishes at equilibrium due to the chirality balance.

Putting everything together, we conclude that only the second term of Eq.~\eqref{eq-currentexplicit} contributes to $\vb j^{(0)}$. 
Thus, the $\tau^0$-scaling response can be written as 
\begin{align}
    \chi^{(0)}_{abc}=& 
    \int[\dd{\vb k}]\left( F_{bc}v^n_a-F_{ac}v^n_b+\varepsilon_{abd}\mathcal M_c^n \Omega_d \right)f'
    \label{eq-chi0}.
\end{align}
This is one of the central results of this section, so we have several comments in order. 
First, $\chi_{abc}^{(0)}$ is anti-symmetric for its first two indices, revealing its dissipationless nature. 
Second, each term in the above equation is purely determined by the band structure and will not be affected by the density or amplitude of the impurities. 
In addition, both $F$ and $\mathcal M$ have orbital and spin degrees of freedom. 
A nonzero spin component requires finite SOC. Otherwise, the two spin states are decoupled, and the magnetotransport simply vanishes. 
Meanwhile, whereas the orbital components do not rely on the SOC, they need a special geometry setup. 
For instance, the bilinear current has no orbital contribution in a 2D system with an in-plane magnetic field since the magnetic moment only has a $z$ component. 
However, the orbital contribution should exist with an out-of-plane magnetic field, and so should ASP. 
Hence, we highlight that the AOP and ASP can generate a purely dissipationless ordinary Hall effect. 
Such a response belongs to the conductivity element $\chi^{(0)}_{yxz}$, an intrinsic term of the ordinary Hall effect, which has not been discussed before.

\subsection{The $\tau^1$-scaling current}
Now we discuss the $\tau^1$-scaling current $j^{(1)}$, which has two distinct properties compared with the $\tau^0$-scaling current. 
First, $j^{(1)}$ requires breaking of time-reversal symmetry, so it only survives in magnetic systems. 
Second, it must involve the nonequilibrium distribution of the Fermi-Dirac function. 
Note that the nonequilibrium distribution is induced by the external fields, and thus it may contain terms linear or even quadratic in the fields. 
Regarding each case, we need the group velocity to be accurate to the linear order of the $E$ field or $B$ field. 
Based on the above considerations, we conclude that only the first and third terms in Eq.~\eqref{eq-currentexplicit} contribute to $j^{(1)}$. 
After some algebra, the conductivity can be represented by
\begin{align}
    \chi_{abc}^{(1)}
    =& -\tau  \int[\dd{\vb k}]\Bigl[
        (\partial_b v^n_{a})\mathcal M_c^n
        -(\partial_a \mathcal M_c^n) v^n_{b}
        \nonumber\\&
        + v^n_{d}\Omega^n_{d}v^n_{b} \delta_{ac}
        - v^n_{a}\Omega^n_{c}v^n_{b}
        + v^n_{a}\Omega^n_{d}v^n_{d}\delta_{bc}
    \Bigr]f'.
    \label{eq-chi1}
\end{align}
Note that the Lorentz force does not modify the local distribution $(\vb v^n\times \vb B)\cdot \partial_{\vb k}f_0=0$, because it only causes the electron to undergo cyclotron motions. 

We can make several observations regarding the above results. 
First, after performing integration by part and noting that $v^n_a=\partial_a \varepsilon_n$, one observes that $\chi^{(1)}$ is symmetric among the first two indices $a$ and $b$, and vanishes in a time-reversal invariant system. 
Second, the first two terms of Eq.~\eqref{eq-chi1} arise from the energy correction due to the $B$ field, which contains both the orbital and spin degrees of freedom. 
Moreover, the third term of Eq.~\eqref{eq-chi1} originates from the flux strength with the magnetic field in the momentum space ($\Omega$). 
The direction of the current is parallel or antiparallel to the $B$ field, depending on the chirality of the electrons. 
Finally, the last two terms of Eq.~\eqref{eq-chi1} come from the nonequilibrium distribution induced by the ``Lorentz-like force'' with a form $\vb (\vb E\times \boldsymbol\Omega)\times \vb B$, where the velocity has been replaced by the anomalous velocity.
Specifically, the fourth term can be treated as an anisotropic magnetoresistance,
while the fifth term exhibits the chiral anomaly of the axion electrodynamics, which only exists in a condition that the two external fields point in the same direction. 
The last three terms exist only in a three-dimensional system and have orbital components.

We note that several works on the PHE have discussed this $\tau^1$-scaling magnetotransport effect~\cite{Ma2019,Zyuzin2020,Battilomo2021,li_Planar_2023a,wang_Fieldinduced_2023}, and Eq.~\eqref{eq-chi1} recovers the previous result under the relaxation time approximation.
In particular, we note that Eq.~\eqref{eq-chi1} can produce a genuine Hall effect because it not only has transverse components but also exhibits an anisotropy between the $a$ and $b$ indices. 
Meanwhile, \citet{Nandy2017a} studied another $\tau$-linear magnetotransport effect with an in-plane magnetic field. 
In particular, they focused on terms quadratic in the $B$ field, so the effect described in their work can appear in non-magnetic systems.

\subsection{The $\tau^2$-scaling current}
Finally, we discuss the $\tau^2$-scaling current $j^{(2)}$, which is a Drude-like response. 
Since the nonequilibrium distribution in Eq.~\eqref{eq-fdis} of the $\tau^2$ term already contains two field corrections as $EB$ factor, we should only retain the field-independent group velocity
of the individual carriers to obtain $j^{(2)}\sim \tau^2$. 
The resulting conductivity can be simplified as
\begin{eqnarray}
    \chi_{abc}^{(2)}&=&
    \tau^2  \varepsilon_{dec}\int[\dd{\vb k}] v_a (\partial_b v_e)v_df'.
    \label{eq-chi2}
\end{eqnarray}
This conductivity contains no interband processes and can be derived within the first-order semi-classical theory, as demonstrated in previous studies~\cite{Xiao2010}. 
It is also $\mathcal T$-even, which can be verified by applying a time-reversal operator to the right-hand side of Eq.~\eqref{eq-chi2}. 
Moreover, it does not have individual spin components, so it does not vanish without SOC.
With a magnetic field applied in the $z$ direction, Eq.~\eqref{eq-chi2} recovers the ordinary Hall effect from the Lorentz force. 
By using $\rho_{xy}\sim \sigma_{xy}/\sigma_{xx}^2$, we can show that the Hall resistivity is independent of the relaxation time $\tau$.

\section{Application to a two-dimensional Dirac model \label{Section:2D}}
We now apply the above general theory to a tilt Weyl point model in 2D, which is a widely studied toy model and has various material representations, such as graphene~\cite{castroneto_Electronic_2009} and surface states of topological insulators~\cite{hasan_Colloquium_2010,qi_Topological_2011}. 
The $\vb k\cdot \vb p$ Hamiltonian can be written as
\begin{eqnarray}
    H= wk_x+ v(k_x s_2+k_y s_1)+\Delta s_3,
    \label{eq-H2d}
\end{eqnarray}
where $s_i$'s are Pauli matrices acting on the spin space, $\Delta$ represents a $\mathcal T$-breaking term, which appears as an exchange field along the $z$ direction and may originate physically from intrinsic magnetic ordering or magnetic proximity effects. 
For convenience, in the following discussion, we assume the mode parameters $w$, $v$, and $\Delta$ to be positive unless specified otherwise.

Due to the presence of the tilt term, the model in Eq.~\eqref{eq-H2d} does not have a rotational symmetry. 
Instead, it only possesses a $\mathsf M_x \mathcal T$ symmetry, which dictates distinct conductivity elements for the $\mathcal T$-even and $\mathcal T$-odd response (see Table~\ref{tab-MxT}).
One can easily check the form of their response tensors with Neumann's principle, which leads to different angular dependencies. 
In what follows, we break down the conductivity according to their different dependencies on $\tau$. 

\begin{table}[t]
    \centering
    \begin{tabular}{cc}
        \hline\hline
        $\mathcal T$-even conductivity & $\mathcal T$-odd conductivity
        \\
        $\chi_{xxx}$,
        $\chi_{xyy}$,
        $\chi_{xyz}$,
        &
        $\chi_{xxy}$,
        $\chi_{xxz}$,
        $\chi_{xyx}$,
        \\
        $\chi_{yxy}$,
        $\chi_{yxz}$,
        $\chi_{yyx}$
        &
        $\chi_{yxx}$,
        $\chi_{yyy}$,
        $\chi_{yyz}$
        \\
        \hline \hline
    \end{tabular}
    \caption{The nonzero conductivity elements under $\mathcal M_x\mathcal T$ for the $\mathcal T$-even and $\mathcal T$-odd conductivities.
        Note that the first two indices of $\chi^{(0)}$ ($\chi^{(1)}$) are anti-symmetric (symmetric).}
    \label{tab-MxT}
\end{table}

\subsection{$\tau^0$-scaling conductivity}
We first study the $\tau^0$-scaling conductivity. 
Two geometric quantities, AOP and ASP, are essential for this quantity, because they establish a relation between the Berry connection and the magnetic field, revealing crucial geometric properties of the band structure. 

\begin{figure}[b]
  \includegraphics[width=0.48\textwidth]{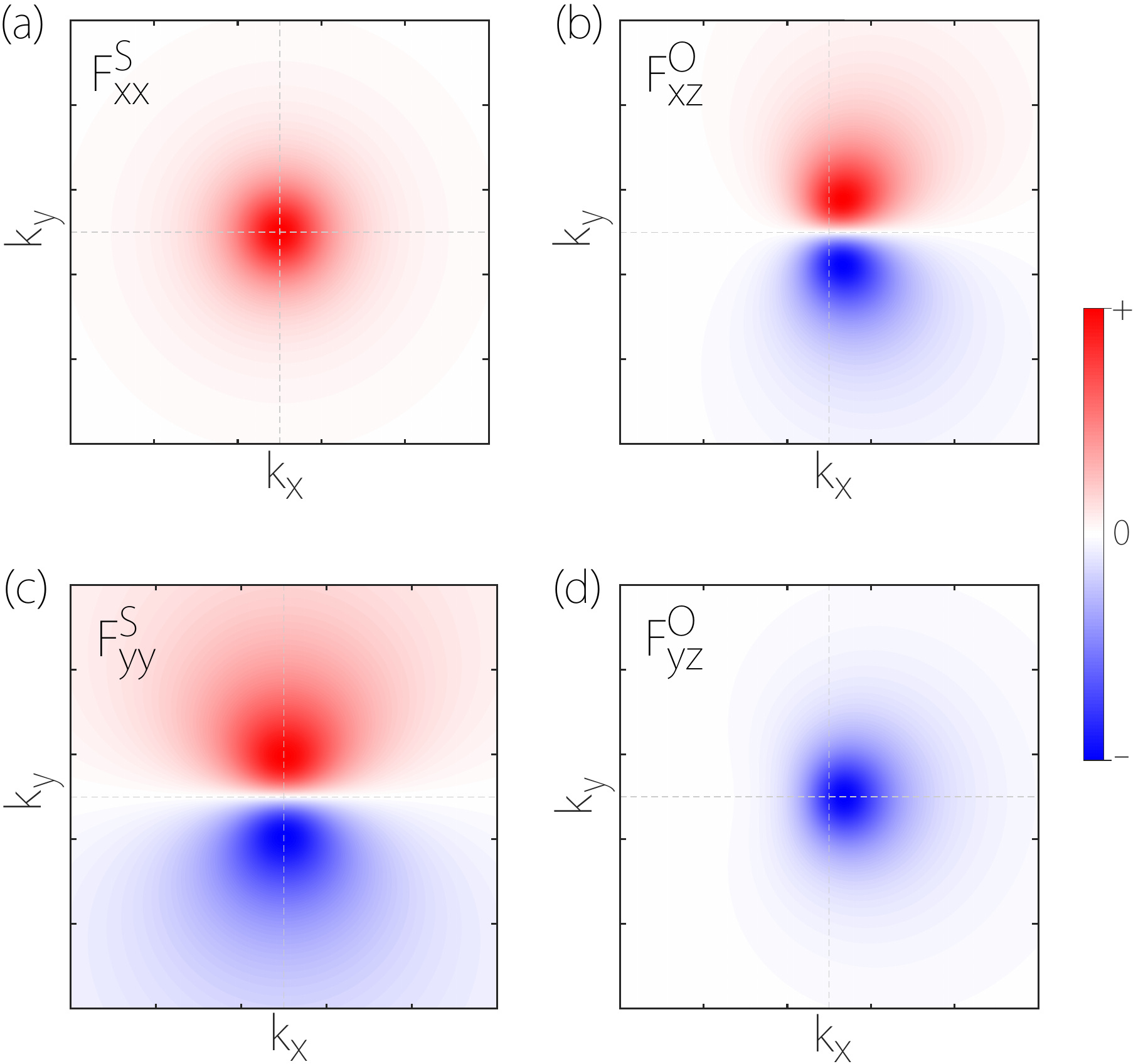}
    \caption{Distribution of $F^{S/O}$ tensor elements in $k$ space. We set $v = \SI{1e6}{m/s}$, $\Delta = \SI{40}{meV}$, $g = 20$, and $w/v = 0.5$. Both ASP and AOP are pronounced near the Dirac point. ASP is centered at $\bm{k}=0$, while the center of AOP is shifted from $k_x=0$ due to the existence of $w$.}
  \label{fig:F}
\end{figure}

Since the tilt term is an overall $k$-dependent energy shift for both energy bands, it does not affect the energy eigenstates.
Therefore, ASP does not depend on $w$; they are isotropic in $k_x$ and $k_y$ directions. 
In contrast, $w$ enters AOP through the magnetic moment.
For the lower band of our two-band model, the two tensors read as
\begin{align}
    [F^O_{ab}]&=  \frac{e  v^2}{8\hbar\varepsilon^5}\mqty[0 & 0& -{v^2k_y (2w k_x +\varepsilon)} \\ 0 & 0& 2 w \varepsilon^2 +v^2 k_x\varepsilon]
    \label{eq-AOP2d},
    \\
    [F^S_{ab}]&=  \frac{g \mu_B v}{4 \varepsilon^3}\mqty[\Delta & 0 & - k_y  v\\ 0 & -\Delta &  k_x v]
    \label{eq-ASP2d},
\end{align}
where $\varepsilon=\sqrt{v^2k^2+\Delta^2}$ and $k=\sqrt{k_x^2+k_y^2}$.
We note that all nonzero elements of these two tensors are consistent with the symmetry constraint $\mathsf M_x\mathcal T$. 
Moreover, AOP does not have any planar elements since the magnetic moment of a $2$D system must be along the $z$ direction. 
In Fig.~\ref{fig:F}, we plot the distribution of AOP and ASP tensor elements in the $k$ space. 
All elements are mainly concentrated in a region around $\vb k=0$, where the local energy gap takes its minimum.
Due to the existence of $w$, AOP's symmetry centers/lines are shifted from $k_x=0$.

With this quantities in mind, the conductivity $\chi^{(0)}$ can be evaluated by using Eq.~\eqref{eq-chi0}.
Here, we recover the factors $\hbar$ and $e$ in the final expressions and use a superscript $S$ and $O$ to denote
the spin and orbital contributions, respectively. 
For the in-plane setup, only spin contribution exists, and the nonzero
element is
\begin{align}
    \chi^{(0),S}_{yxy}&= -\chi^{(0),S}_{xyy}= \frac{e^2}{\hbar}
    \frac{g \mu_B w \Delta \mu v^2}{8\pi (\mu^2v^2+\Delta^2 w^2)^{3/2}}.
    \label{eq-chi0syxy}
\end{align}
We note that our results apply to both type-I and type-II weyl points.
This conductivity is linear in $w$ and $\Delta$, confirming our symmetry analysis that any out-of-plane rotation axis would render all components of $\chi_{abc}$ to be zero. 
Moreover, due to the fermion doubling theorem, any non-magnetic system cannot have one single Weyl point. 
In the simplest case, the system should contain another Weyl point connected by the time-reversal symmetry to the one we wrote in Eq.~\eqref{eq-H2d}, which has opposite values of $w$ and $\Delta$. 
The two points will have the same contribution instead of opposite signs to the bilinear magnetotransport effect. 
One can observe that $\chi^{(0),S}$ dominates near the band gap and decreases as $\sim 1/\mu^2$ with increasing $\mu$, indicating its geometric origin.

\begin{figure}[t]
  \includegraphics[width=0.48\textwidth]{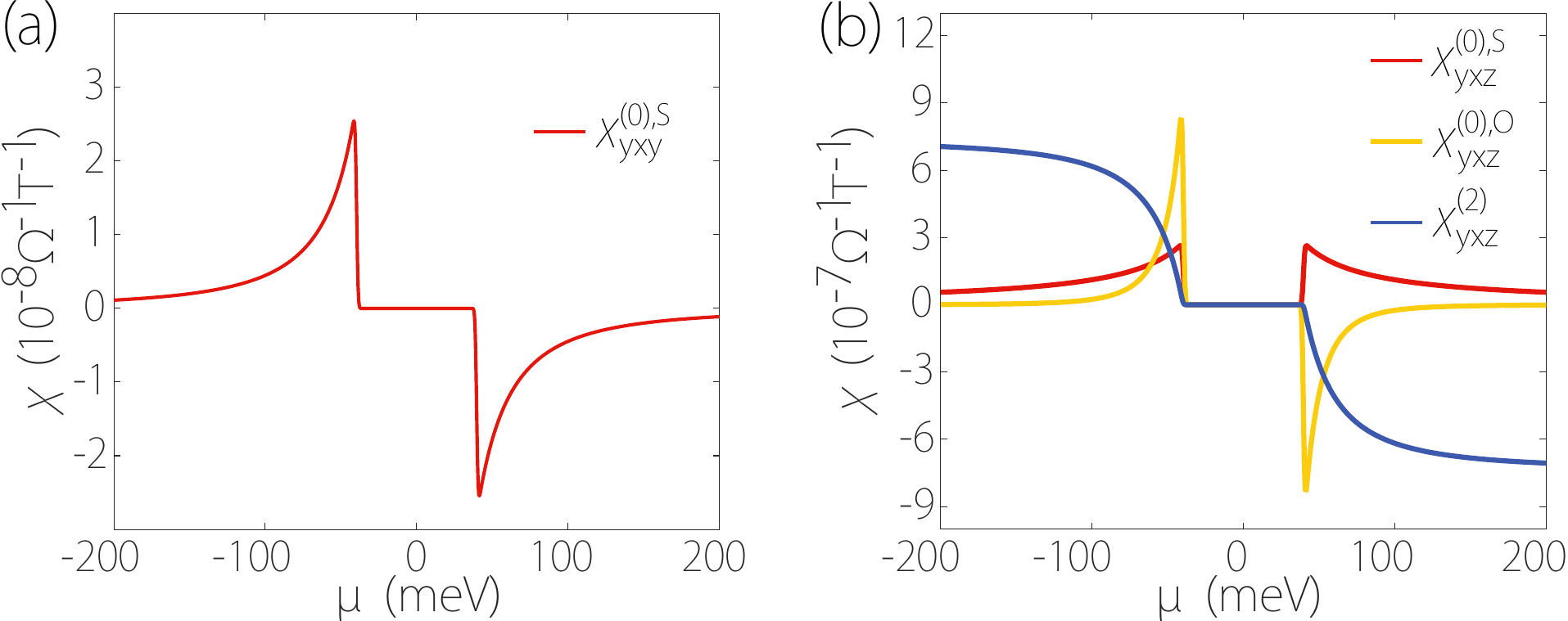}
    \caption{ Conductivities of PHE (a) $\chi_{yxy}$ and ordinary Hall (b) $\chi_{yxz}$ effect as a function of $\mu$.
        The $\tau^0$-scaling conductivities take their peaks near the band edges, where the energy gap takes its minimum. While the $\tau^2$-scaling conductivity increases with the chemical
        potential goes away from the band edges.
        In our calculation, we choose typical parameters as $v/\hbar = \SI{5e5}{m/s}$,
    $w/v = 0.1$, $\Delta =\SI{40}{meV}$, $g = 20$, and $\tau = \SI{10}{fs}$.}
  \label{fig:yxz0yxy0}
\end{figure}

When the $B$ field is applied in the $z$ direction, both orbital and spin contributions exist, given by 
\begin{align}
    \chi^{(0),S}_{yxz}&= \frac{e^2}{\hbar}
    \frac{g\mu_B \mu^2 v^3}{8\pi (\mu^2 v^2+\Delta^2 w^2)^{3/2}},
    \label{eq-chi0Syxz}
    \\
    \chi^{(0),O}_{yxz}&= \frac{e^2}{\hbar}
    \frac{e \mu v^3 \left[v^2 w^2 \left(\Delta^2+\mu^2\right)-2 \Delta^2 v^4+\Delta^2 w^4\right]}{16 \pi \hbar \left(\Delta^2 w^2+\mu^2 v^2\right)^{5/2}}
    \label{eq-chi0Oyxz}.
\end{align}
These two conductivities are classified as ordinary Hall conductivity, although they stem from ASP and AOP but not the Lorentz force.
Moreover, the spin and orbital contributions are pronounced near the band edges, consistent with the geometric quantities shown in Fig.~\ref{fig:F}.
For the homogeneous case with $w=0$, the orbital contributions decreases as $\mu^{-4}$ while the spin contribution decreases as $\mu^{-1}$ with increasing $\mu$.
We plot $\chi^{(0)}_{yxz}$ and $\chi^{(0)}_{yxy}$ as a function of $\mu$ in Fig.~\ref{fig:yxz0yxy0}(a) and \ref{fig:yxz0yxy0}(b).
One can observe that the orbital contribution dominates around the band edges but decreases quickly with a chemical potential going away from the band edges.
The spin contribution requires a finite SOC, as we mentioned above. 
In fact, there is a simple relation between the orbital and spin contributions $\chi^{(0),S}/\chi^{(0),O}={g\mu_B \hbar \mu^3}/{(ev^2 \Delta^2)}$.
A large $g$ factor will enhance the spin contribution, and the orbital contribution requires a finite $\Delta$.

\subsection{$\tau^1$-scaling conductivity}

The $\tau^1$-scaling current can only exist in a time-reversal breaking system, so it is also called a $\mathcal T$-odd response.
In the presence of $\mathsf M_x \mathcal T$, the ordinary Hall responses $\chi_{xyz}^{(1)}$ and $\chi_{yxz}^{(1)}$ are forbiden, as shown in Table~\ref{tab-MxT}.
Moreover, the orbital part can not provide any planar Hall signals.
From our calculation, the spin part also vanishes when $a\ne b$.
Therefore, no Hall component of $\chi^{(1)}$ exists; 
this current can only contribute to the longitudinal magnetoresistance.

We now present the analytical results for the $\tau^1$-scaling responses in the zero-temperature limit. 
When the applied $B$ field lies in the transport plane, only the spin component has a nonzero element, given by 
\begin{align}
    \chi^{(1),S}_{yyy}= \frac{e^2 \tau}{\hbar^2} \frac{g\mu_B}{4\pi w} \left( \frac{1}{\sqrt{v^2-w^2}}-\frac{1}{\sqrt{v^2+\Delta^2 w^2/\mu^2}} \right),
    \label{eq-chi2Syyy}
\end{align}
which is an odd function of $w$ and an even function of $\Delta$.
In Eq.~\eqref{eq-H2d}, the time-reversal symmetry is broken by $w$ and $\Delta$, consistent with the $\mathcal T$-odd nature of $\chi^{(1)}$. 
In fact, the above result is the chiral magnetoresistance effect from the nonzero $\vb E\cdot \vb B$. 

\begin{figure}[t]
  \includegraphics[width=0.48\textwidth]{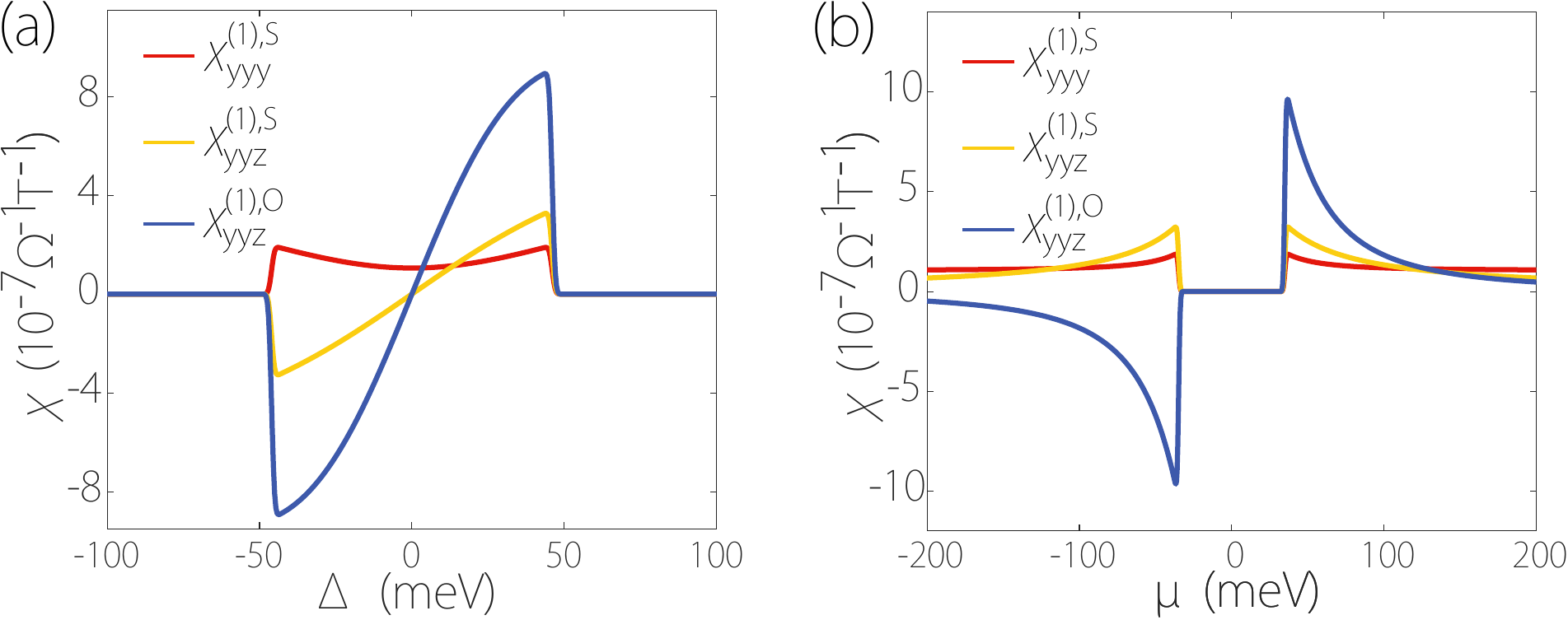}
  \caption{ $\chi^{(1)}_{yyy}$ and $\chi^{(1)}_{yyz}$ as a function of (a)$\Delta$ and (b)$\mu$. We set $v/\hbar = 5\times10^5\text{m/s}$, $w/v = 0.5$, $g = 20$, and $\tau = \SI{10}{fs}$. In (a), we set $\mu = -\SI{40}{meV}$, which implies that only the valence band contributes to the response. The components of the response tensor have different parity with respect to time. In (b), we set $\Delta = \SI{40}{meV}$, the orbital contribution is an odd function of the $\mu$, while the spin contribution is the opposite.}
  \label{fig:yyz1}
\end{figure}

When the magnetic field is parallel (antiparallel) to the electric field, it leads to a negative (positive) magnetoresistance, as shown in Fig.~\ref{fig:yyz1}. 
In this case, $\chi^{(1),S}_{yyz}$ is an odd function of $\Delta$:
\begin{align}
    \chi^{(1),S}_{yyz}&= \frac{e^2}{\hbar} \frac{g\mu_B\tau}{4\pi \hbar} \frac{v \Delta}{(\mu^2v^2+\Delta^2w^2)^{1/2}}
    \label{eq-chi1yyzs}.
\end{align}
As mentioned above, the orbital term only remains finite when the $B$ field is applied in $z$ direction, and the corresponding element reads as 
\begin{align}
    \chi^{(1),O}_{yyz}&= 
     \frac{e^2}{\hbar} \frac{e\tau}{4\pi \hbar^2} \frac{\Delta \mu v^5}{(\mu^2v^2+\Delta^2w^2)^{3/2}}. 
    \label{eq-chi1yyz}
\end{align}
All these elements also reveal their band-geometric nature, as their peaks appear near the band edges. 
Fig.~\ref{fig:yyz1} shows $\chi^{(1),i}_{yyz}$ and $\chi^{(1),i}_{yyy}$ as a function of the chemical potential $\mu$ and energy gap $\Delta$, with $i=S,O$.
Apart from the different scaling relations with the chemical potential, we draw the readers' attention to the different signs of these conductivities. 
In particular, while $\chi^{(1),S}_{yyy}$ and $\chi_{yyz}^{(1),S}$ are always positive, $\chi^{(1),O}_{yyz}$ changes sign as one moves $\mu$ from the conduction band to the valence band.

\subsection{$\tau^2$-scaling conductivity}
The $\tau^2$-scaling contribution only has an orbital part. 
It follows that there is no signal with the in-plane setup. 
As we discussed above, this term is the classical Drude-like contribution.
When the $B$ field lies in $z$ direction, the transverse component is the well-known Hall effect, and the conductivity is represented by 
\begin{align}
    \chi^{(2)}_{yxz}&= - \frac{e^3\tau^2}{\hbar^4}
    \frac{\mu^3 v^5+ \mu\Delta^2 v^3(w^2-v^2)}{4\pi(\mu^2 v^2+\Delta^2 w^2)^{3/2}}
    \label{eq-chi2yxz}.
\end{align}
It should be anti-symmetric for the first two indices, as has been proved by our calculations.

It is interesting to compare $\chi^{(0)}_{yxz}$ and $\chi^{(2)}_{yxz}$.
Both contributions come from the same setup but may originate from different mechanisms.
The Drude-like conductivity $\chi^{(2)}_{yxz}$ does not arise from geometric quantities that are enhanced near small energy gaps. 
In fact, when $w=\Delta=0$, $\chi^{(2)}_{yxz}$ only depends on the Fermi velocity $v$ and the relaxation time $\tau$.
In contrast, $\chi^{(0)}$ arises from the anomalous velocity induced by the Berry curvature, so it generally strongly depends on the geometry of the band structure. 
In our model, these two quantities are related by 
\begin{align}
    \frac{\chi^{(2)}_{yxz}}{\chi^{(0),O}_{yxz}}
    &=  \frac{2\mu^2\tau^2(\Delta^2-\mu^2)}{\hbar^2\Delta^2},
    \\
    \frac{\chi^{(2)}_{yxz}}{\chi^{(0),S}_{yxz}}&= -\frac{2 e \tau^2 \left[\Delta^2 \left(w^2-v^2\right)+\mu^2 v^2\right]}{g \mu \mu_B}.
    \label{eq-compare02}
\end{align}
As expected, the intrinsic contribution dominates near the small-gap regime $\mu \sim \pm \Delta$.
We present the comparison of the $\tau^0$ and $\tau^{2}$ conductivities in Fig.~\ref{fig:yxz0yxy0}(b).

\section{$\mathcal{PT}$ symmetry in three-dimension \label{Section:3D}}
As another example, we apply our theory to a spinfull $\mathcal{PT}$-symmetric model. 
The existence of $\mathcal{PT}$ suppresses the anomalous Hall effect and the second-order Hall effect from the Berry curvature dipole. 
As a result, the leading Hall signal should be the nonlinear anomalous Hall effect induced by the Berry connection polarizability and the bilinear current studies in this work. 
The model we study is given by 
\begin{align}
    H=w_1k_x+w_3k_z+vk_x\sigma_3+v k_y\sigma_2 s_1+vk_z\sigma_1+\Delta\sigma_2s_3,
    \label{eq-fourbandm}
\end{align}
where $\sigma_i$'s are Pauli matrices acting on the orbital degrees of freedom.
In this model, we assume the Fermi velocities are identical along the three directions. 
The general anisotropic cases can be accommodated by rescaling the wave vectors $k_a$. 

There are some remarks before processing. 
First, the $\mathcal{PT} =i s_2\mathcal K$ symmetry ensures a double degeneracy at every $\vb k$ point. 
Therefore, the calculation must use the non-Abelian framework to handle the coherent degeneracy bands.
Second, the pair of degenerate bands have identical group velocities, opposite Berry curvatures, and opposite magnetic moments due to the $\mathcal{PT}$ symmetry. 
These properties ensure a vanishing $\tau^1$-scaling current according to Eq.~\eqref{eq-chi1}. 
Therefore, there are only $\mathcal T$-even contributions.
Finally, to guarantee the planar Hall effect by AOP, it is necessary to have a tilted term along the $z$ direction, which may be induced by a strain~\cite{pereira_Tightbinding_2009} or an external field. 

\begin{figure}[t]
  \includegraphics[width=0.48\textwidth]{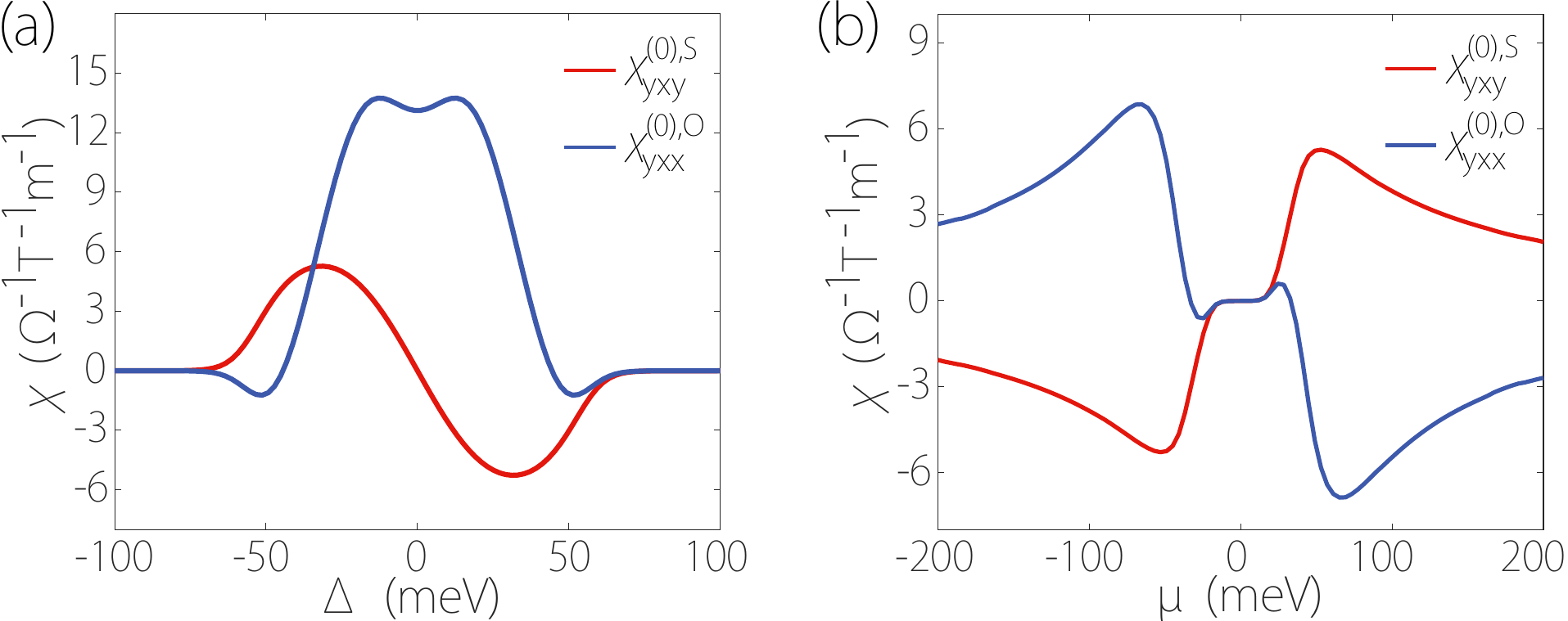}
  \caption{ $\chi^{(0),S}_{yxy}$ and $\chi^{(0),O}_{yxx}$ as a function of (a) $\Delta$ and (b) $\mu$. We set $v/\hbar = \SI{5e5}{m/s}$, $w_1/v=w_3/v = 0.5$, $\tau = \SI{10}{fs}$, $g = 20$, and $T = \SI{30}{K}$. In (a), we set $\mu = \SI{-40}{meV}$. In (b), we set $\Delta = \SI{40}{meV}$.}
  \label{fig:yxxyxy3SO}
\end{figure}

In Fig.~\ref{fig:yxxyxy3SO}, we plot the numerical results for the orbital and spin planar Hall response elements. 
One can see that both quantities exhibit a similar qualitative behavior to the $\chi_{yxz}^{(0),O/S}$ in Fig.~\ref{fig:yxz0yxy0}(b), as they peak around the small gap edges. 
Such a feature reveals that $\chi^{(0),S}_{yxy}$ and $\chi^{(0),O}_{yxx}$ rely on band geometric quantities. 

In addition, we note that the orbital and spin contributions can be easily separated by their different qualitative features. 
Specifically, the introduction of the tilt $w_3k_z$ along the $z$ direction does not affect ASP, and hence just slightly modified $\chi^{(0),S}$. 
In contrast, $\chi^{(0),O}$ strongly depends on $w_3$. 
Besides, from Fig.~\ref{fig:yxxyxy3SO}(a), one can immediately observe that $\chi^{(0),S}$ is an odd function of $\Delta$, whereas $\chi^{(0),O}$ is an even function. 
Using this property, one can separate them by analyzing the response under the reversal of the magnetic ordering.

\section{Discussion and conclusion \label{Section:Discussion}}
We note that the Hall voltage induced by these quantities is large enough to be measured in the experiment. 
Specifically, the Hall voltage can be estimated by $V_H\sim \chi EB L/\sigma_{xx}$, where $L$ is the width of the sample and $\sigma_{xx}$ is the longitudinal conductivity. 
Taking the 2D model as an example, where we found that $\chi\sim \SI{1e-7}{\ohm^{-1}T^{-1}}$. 
Considering that $E=\SI{1e4}{V/m}$, $B=\SI{1}{T}$, $L=\SI{10}{\mu m}$, and a typical $2$D conductivity $\sigma_{xx}=\SI{1e-4}{\ohm}$, we can obtain a Hall voltage about $\SI{100}{\mu V}$, which is certainly within the experimental capabilities~\cite{Kang2019a,Lai2021c}. 
Moreover, the response in an actual material may be further enhanced by specific features in the band structure. 
For example, the intrinsic term should be pronounced in the semimetal with various near-degeneracy bands and strong SOC. 
For a bulk topological semimetal $\mathrm{SrAs_3}$~\cite{Wang2024a}, the intrinsic spin contribution is about $\SI{4}{\ohm^{-1}m^{-1}T^{-1}}$, while the orbital contribution is about $\SI{143}{\ohm^{-1}m^{-1}T^{-1}}$, much larger than our model calculations. 
Recent expriments in heterodimensional $\mathrm{VS_2}$\textendash$\mathrm{VS}$ superlattice reported a conductivity as large as $\SI{800}{\ohm^{-1}m^{-1}T^{-1}}$~\cite{Zhou2022a}.

We also note that the contributions from different $\tau$ scaling can be extracted with the information of the longitudinal conductivity, which is known to vary linearly with $\tau$. 
A commonly used method is to plot the response signals versus $\sigma_{xx}$~\cite{Kang2019a,Lai2021c} and fit the curve with some scaling law. 
For example, in our case, we should use a function like $y=c_0+c_1x+c_2x^2$ to fit the various $\tau$ scaling components, where $c_i$ denotes the amplitudes of each $\tau^i$-scaling contributions and are independent of $\tau$. 

Finally, we note that the present work is based on the relaxation time approximation and has ignored many possible disorder-induced effects~\cite{Nagaosa2010a}. 
For disorder from multiple sources, the scaling law may be much more complicated~\cite{Hou2015}, which is beyond the scope of this work.


In conclusion, we have presented a unified theory for the bilinear current response, scaling both linearly with $E$ and $B$ fields. 
It includes the classical Drude-like response as well as the field-induced quantities.
Depending on the driving and measurement setup, this current can be classified into ordinary Hall effect, PHE, and magnetoresistance. 
We demonstrated that the novel quantum geometric quantities AOP and ASP play essential roles in the ordinary Hall effect and the dissipationless PHE, which are particularly pronounced in cases where near-degenerate bands appear near the Fermi surface.

\section*{Acknowledgement}
This work is supported by the Research Grants Council of Hong Kong (Grants No. CityU 21304720, No. CityU 11300421, No. CityU 11304823, and No. C7012-21G) and City University of Hong Kong (Projects No. 9610428 and No. 7005938).

\bibliography{ref1}
\bibliographystyle{apsrev4-2}

\appendix
\section{Conductivities for the $3$D model with different scaling relation to $\tau$}
\begin{figure}[ht]
  \includegraphics[width=0.48\textwidth]{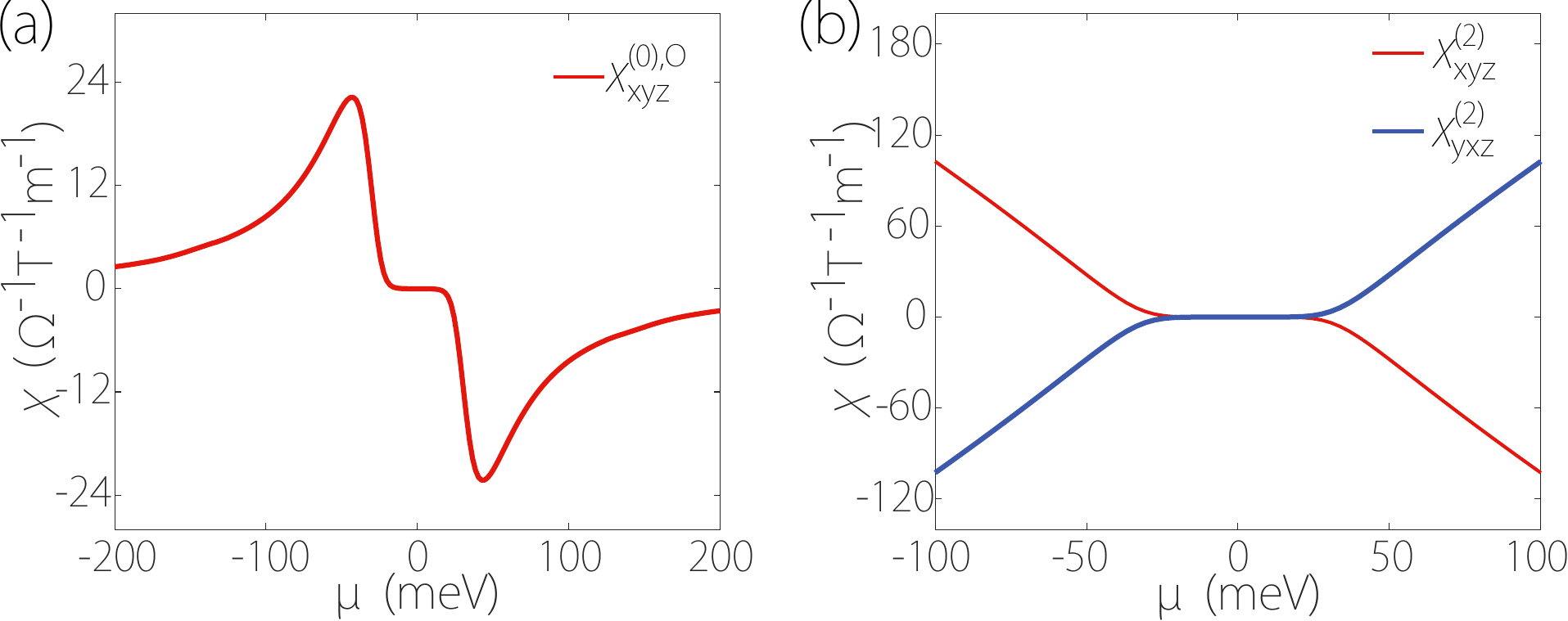}
  \caption{The $\chi^{(0),O}_{xyz}$ as a function of $\mu$. The $\tau^0$-scaling conductivities take their peaks near the band edges. 
        We set $v/\hbar = \SI{5e5}{m/s}$,
    $w/v = 0.5$, $\Delta =\SI{40}{meV}$, $g = 20$, and $T =\SI{30}{K}$. As the Fermi surface moves away from the energy gap, the second-order signal gradually increases.}
  \label{fig:xyz0Oxyzyxz2}
\end{figure}
For $\tau^0$-scaling conductivity, the result of $\chi^{(0)}$ involves two geometric quantities ASP and AOP. The properties of the three-dimensional results have many similarities with the two-dimensional ones. For the in-plane response, there are both spin and orbital contributions present.
However, if there is an out-of-plane component in the index, the spin contribution vanishes, leaving only the orbital part, shown in Fig.~\ref{fig:xyz0Oxyzyxz2}.

For $\tau^1$-scaling conductivity, the bands in the degeneracy subset share the same group velocity but opposite Berry curvature and magnetic moments, due to the $\mathcal{PT}$ symmetry. One can easily check that the two degeneracy bands
have opposite contributions to $\chi^{(1)}$ from Eq.~\eqref{eq-chi1}.
Hence, $\chi^{(1)}$ vanishes under a $\mathcal{PT}$ symmetry.

For $\tau^2$-scaling conductivity, we present the $\chi_{xyz}^{(2)}$ and $\chi_{yxz}^{(2)}$ components in Fig.~\ref{fig:xyz0Oxyzyxz2}. as examples here. The $\tau^2$ response increases as the Fermi surface moves further away from the energy gap, similar to the scenario we encountered in the 2D system.

\end{document}